\documentclass[preprint,12 pt]{article}
\usepackage{arxiv}

\usepackage{cite}

\usepackage{float}
\usepackage{amsmath,amssymb,amsfonts}
\usepackage{algorithmic}
\usepackage{graphicx}
\usepackage{textcomp}
\usepackage{color,soul}
\usepackage{flushend}
\usepackage{cite}
\usepackage{mathtools}
\usepackage{array,multirow}
\usepackage[ruled,linesnumbered]{algorithm2e}

\usepackage{amsmath}
\usepackage{amsthm}
\usepackage{graphicx}          % Include this line if your 
\usepackage{amsmath} % assumes amsmath package installed
\usepackage{amssymb}  % assumes amsmath package installed
\usepackage{color,soul}
\usepackage{enumitem}
\usepackage{cite}
\usepackage{cleveref}
\usepackage[ruled,linesnumbered]{algorithm2e}
\usepackage{dblfloatfix}

\title{Secure Control of Connected and Autonomous Electrified Vehicles Under Adversarial Cyber-Attacks}
\author{Shashank Dhananjay Vyas \thanks{During his part of this work, S. D. Vyas was with The Pennsylvania State University.}\\
Department of Mechanical Engineering\\ The Pennsylvania State University\\ University Park, Pennsylvania 16802, USA\\E-mail: sbv5192@psu.edu.\\
\And
Satadru Dey\\
Department of Mechanical Engineering\\ The Pennsylvania State University\\ University Park, Pennsylvania 16802, USA\\E-mail: skd5685@psu.edu.\\}

\begin{document}

\maketitle

\begin{abstract}
Connected and Autonomous Electrified Vehicles (CAEV) is the solution to the future smart mobility having benefits of efficient traffic flow and cleaner environmental impact. Although CAEV has advantages they are still susceptible to adversarial cyber attacks due to their autonomous electric operation and the involved connectivity. To alleviate this issue, we propose a secure control architecture of CAEV. Particularly, we design an additional control input using Reinforcement Learning (RL) to be applied to the vehicle powertrain along with the input commanded by the battery. We present simulation case studies to demonstrate the potential of the proposed approach in keeping the CAEV platoon operating safely without collisions by curbing the effect of adversarial attacks. 
\end{abstract}

% \begin{IEEEkeywords}
% Connected and Autonomous Vehicles, Electric Vehicles, Cyber-Physical Security.
% \end{IEEEkeywords}

\section{Introduction}
\label{sec:introduction}
Connected and Autonomous Electrified Vehicles (CAEV) is a platoon of autonomous vehicles (AV) which are battery powered and there is a communication network present which enables exchange of information between the vehicles in the platoon. Integration of these three components namely, autonomy, connectivity and electrification, enhances road safety, reduces traffic congestion and improves fuel efficiency. This results in increase in traffic throughput, supports the concept of shared autonomous fleets using clean energy and reduction in carbon emissions \cite{vaidya2019connected,lu2019energy}. One of the early works bringing Connected and Autonomous Vehicles (CAV) and Electric Vehicle (EV) together is \cite{alkheir2019connected}. Due to these various advantages of CAEV, they are considered to be the future of smart mobility.

\subsection{Literature Review}

There have been some works in the literature focussed on control of CAEV. In many of the research articles, the area targeted is energy utilization in CAEVs. For example, an energy saving eco-controller is presented in \cite{bertoni2017adaptive} to minimize the energy consumption in CAEV. An adaptive cruise control is proposed in \cite{lu2019energy} for energy eficient operation of CAEV. A Nonlinear Model Predictive Controller is developed in \cite{coppola2022eco} for optimal energy comsumption in CAEV. Some works are focussed on other CAEV control objectives. Such as in \cite{guo2020adaptive} a nonlinear adaptive control for CAEV on curved roads is proposed for supervising coupled lateral and longitudinal dynamics.

Although there are various advantages of CAEV which existing works have tried to maximize, CAEV are still vulnerable to cyber attacks due to the information transfer involved through communication network and sensors. Existing works are mostly focussed on the issue of safety and security of CAV, EV and batteries. A resilient control strategy to mitigate the effect of Denial-of-Service (DoS) attacks in CACC equipped vehicular platoons is proposed in \cite{biron2017resilient}. A cyberattack detection algorithm consisting of unified Vehicle-toVehicle (V2V) and Vehicle-to-Infrastructure (V2I) communication and a V2I cyber attack isolation scheme for CAV with changing driving conditions is proposed in \cite{ghosh2024cyberattack}. A zero-trust architecture is proposed in \cite{anderson2023zero} to intrinsically protect the individual sensors of CAV and the internal vehicle networks using the CAN bus. A survey of cyber security as applicable to the CAV environment is presented in \cite{sun2021survey}. A security assessment of the EV charging infrastructure is presented in \cite{antoun2020detailed} highlighting and categorizing the cyber threats. A machine learning-based controller which makes use of an adaptive boosting-based attack detector and a Long Short Term Memory (LSTM) attack estimator for attack resilient battery systems is proposed in \cite{srinath2024machine}. In \cite{11072487}, a threat modeling framework has been proposed for battery systems. Still, a few works have tried to address the problem of safety in CAEV. A game theory based longitudinal control framework integrating economical energy use and safety is proposed in \cite{cheng2020longitudinal}. An analytical state-constrained optimal control problem is formulated in \cite{han2018safe} to minimize CAEV energy usage while keeping safety and following road limits. Although these control works consider the safety of CAEV but controllers focussed on addressing cyber security of CAEV are less prominent in the literature. Hence, secure control of CAEV is a relatively underexplored area.

\begin{figure*}
\centering
\includegraphics[width=\textwidth]{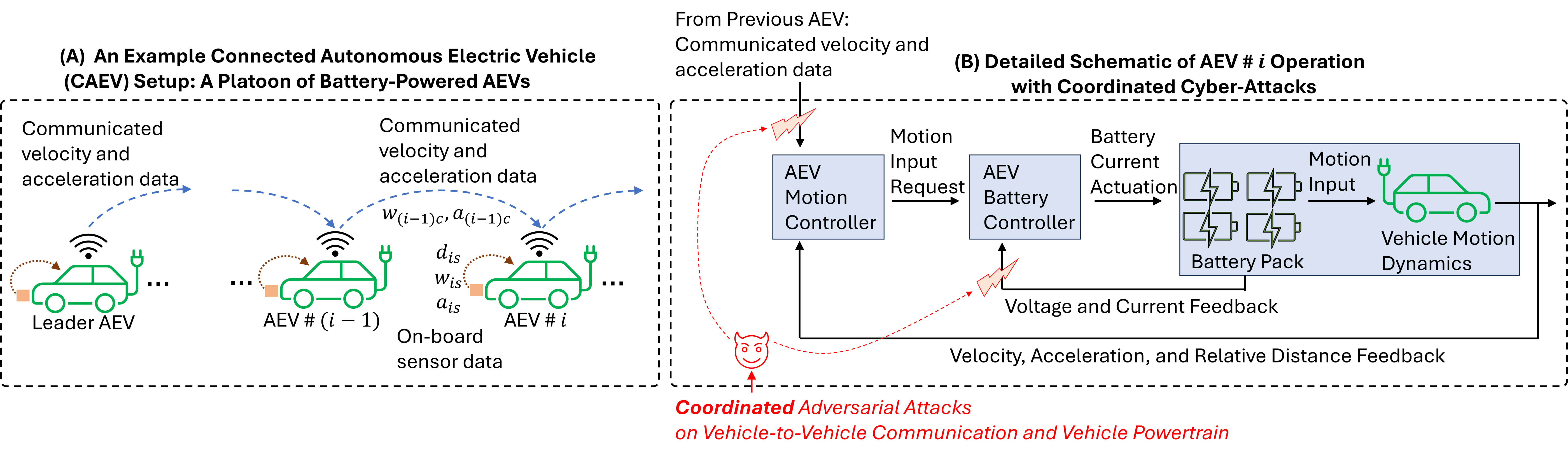}
\caption{(A) An example schematic of a battery-powered platoon under Coordinated Adaptive Cruise Control (CACC) strategy. (B) An expanded view of the local autonomous vehicle controller operating as a part of the vehicle platoon.} 
\label{fig:diagCAEV}
\end{figure*}

% \textcolor{red}{Para 3: Discuss how CAEV cybersecurity if different from CAV or AV security.}

\subsection{Research Gaps and Main Contribution}

CAEV cyber security involves defending against both, attacks occuring due to vulnerability in autonomous electric operation and attacks due to corrupted data in connectivity. There are existing works to counter the unwanted effects occuring due to these attacks such as faults in batteries and undesirable controller behaviour in CAV. But these works are focussed on addressing issues individually such as battery fault detection \cite{vyas2024energy} and control \cite{vyas2022thermal,vyas2024input}, CAV attack detection \cite{vyas2023physics,10320379,9312453}. Eliminating the effects of combination of attacks is far less common in literature. In this work, we focus on this target area, i.e., desigining the secure control framework which will mitigate the effect of coordinated attacks on autonomous battery current snesor and communicated acceleration data.

% \textcolor{red}{Para 4: Start the para with your contribution statement. Then discuss the elements of your approach.}
In light of aforementioned research gap, we propose a secure control architecture for CAEV under the presence of adversarial cyber attacks. We consider the platoon to be operating under a Cooperative Adaptive Cruise Control (CACC) strategy while interacting with the battery present in each vehicle. The adversary is considered to be affecting the current sensor in the battery and the intervehicular communication network with the malicious goal of causing vehicle collision. In order to mitigate the effect of the corrupted current sensor and communicated acceleration data, we develop a defender using Reinforcement Learning (RL) with the objective of maintaining desired standstill intervehicular distance. Specifically, we train a RL agent to give an additional control input to the vehicle powertrain along with the power supplied by the battery under adversarial conditions. We have used Proximal Policy Optimization (PPO) algorithm as the RL agent \cite{schulman2017proximal} in this work. %We refer the reader to \cite{9904958} for a comprehensive review of Deep RL theory and prominent algorithms.

The rest of the paper is organized as follows. In Section II, we discuss the issue of cyber physical security in CAEV mathematically stating the attack vectors. Section III details the secure control framework using the RL agent. In Section IV, we present the results from simulation case studies. Finally, Section V concludes the paper.

\section{Cyber-Physical Security in Connected Autonomous Electric Vehicles (CAEV)}

In this section, we discuss the operational description, security scenario, and modeling of CAEV setup.

\subsection{Operational Description of CAEV}

We consider a homogeneous platoon of CAEV as shown in Fig. \ref{fig:diagCAEV}. The vehicles follow the Cooperative Adaptive Cruise Control (CACC) strategy. The battery inside each vehicle gives power to the vehicle based on the control input decided by the CACC strategy. First, we discuss the high level CACC-based platoon operation as follows. The platoon under consideration consists of a leader-follower structure. Under this formation, the leader vehicle follows a reference velocity profile, $w_{ref}$, and the objective of each of the follower vehicles is to keep a desired intervehicular distance from its preceding vehicle. We denote the states of the any $i^{th}$ vehicle as $[d_i\ w_i\ a_i]^T$ where $w_i$ and $a_i$ are its velocity and acceleration, respectively, and $d_i$ is the distance from its predecessor. Each follower vehicle has on board sensors which are used to measure the following: (i) the intervehicular distance $d_{is}$ from its preceding vehicle, (ii) its velocity $w_{is}$, and (iii) its acceleration $a_{is}$. The subscript `$s$' indicates that it is measurement (e.g. $w_{is}$) and not the actual system state ($w_i$). Along with this, it also receives velocity $w_{(i-1)c}$ and acceleration $a_{(i-1)c}$ information of the preceding vehicle via Dedicated Short Range Communication (DSRC) network. The subscript `$c$' indicates that it is measurement (e.g. $w_{(i-1)c}$) and not the actual system state ($w_{(i-1)}$). Next, we discuss the role of battery in achieving the desired CACC objective inside each vehicle as follows. The local controller in each vehicle computes a desired acceleration input to achieve the CACC objective and sends that input requirement to a input-to-power (u/P) converter which converts the required acceration input $u_{req}$ to a power requirement $P_{req}$ and sends it to the battery. The current controller is a key part of the Battery Management System (BMS) which sends the current to the battery so that the required power demand can be met. The battery delivers the power $P$ through the power-to-input (P/u) converter which converts $P$ to the acceleration input $u$ supplied to the vehicle.

\subsection{Cyber-Attacks in CAEV}
Finally, we discuss the security aspects of the platoon operation of battery powered vehicles. The attacks can occur through various attack surfaces present in the platoon such as the DSRC network, local vehicle state sensors and battery sensors sending data to the BMS. In this work, we consider two attack vectors: (i) attack on the acceleration data $a_{(i-1)c}$ received by the local vehicle controller through intervehicular communication, and (ii) attack on the current data $I$ received by the BMS through current sensor on the battery. The rest of the paper is written from the point of view of the $1^{st}$ follower vehicle, for the sake of brevity we drop the subscript $i$ when referring to the $i^{th}$ vehicle ($i=1$ in this case) and use the term 'ego' vehicle whenever required. Further, we assume that we can measure all the states of vehicle and hence drop the subscript $s$ in the rest of the paper.

\subsection{Modeling of CAEV}
We now discuss the mathematical modeling of the CAEV using the aforementioned CACC strategy with the battery in the control loop. The ego vehicle's motion can be described by the following equations \cite{5531596}
\begin{align}\label{ssveh}
        \dot{d} &= w_c - w, \\
        \dot{w} &= a, \\
        \dot{a} &= -\frac{1}{h} a + \frac{1}{h} u,
\end{align}
where $d$, $w$, and $a$ are the relative distance between the leader and the ego vehicle, velocity of the ego vehicle, and acceleration of the ego vehicle. The coefficient $h$ is the time headway constant and $u$ is the control input received by the powertrain. Under the CACC strategy, the input $u_{req}$ determined by the controller is given by
\begin{align}\label{ureq}
    u_{req} = (k_p e + k_d \dot{e}) + a_c,
\end{align}
where
\begin{align}\label{err}
    e = d - hw - d_r,
\end{align}
is the error of the tracking controller (CACC) with reference distance = $d_r + h w$, $d_r$ is the desired standstill distance between successive vehicles, and $k_p, k_d$ are the controller gains. The $u_{req}$ is sent to the u/P converter which converts the required control input into power requirement $P_{req}$ as
\begin{align}\label{utoP}
    P_{req} = \kappa u_{req}
\end{align}
where $\kappa$ is the gain and the required power $P_{req}$ is sent to the current controller in the BMS. Next, we use the Single Particle Model (SPM) framework to capture battery anode dynamics \cite{santhanagopalan2006online}:
\begin{align}
    &\frac{\partial x_b}{\partial t} = \frac{D}{r^2} \frac{\partial}{\partial r}\left(r^2 \frac{\partial x_b}{\partial r}\right), \frac{\partial x_b(0)}{\partial r} = 0, \frac{\partial x_b(r_a)}{\partial r} = -\gamma_b I,\label{conceqbc} \\
    &V = V_a(x_b) - h(I), \label{volteq}
\end{align}
where $x_b$ is Lithium concentration along anode particle with radius $r_a$ ($r$ being the radial coordinate), $I$ is the current, and $V$ is the terminal voltage of the battery; $D$, $\gamma_b$, $V_a(.)$, $h(.)$ are battery model parameters which can be found in \cite{santhanagopalan2006online,dey2015nonlinear}. Following the finite dimensional approximation used in \cite{dey2015nonlinear}, an Ordinary Differential Equation (ODE) battery model can be formulated in the following form:
\begin{align}\label{ssbatt}
    \dot{c}_a = A_b c_a + B_b I,
\end{align}
Next, we consider that the current controller to be governed by a proportional control law given by
\begin{align}\label{ctrlbatt}
    I = K_b (P_{req} - P_f),
\end{align}
where $P_{req}$ is the power request from the battery and $P_f$ is the power computed from the voltage and current sensor measurements as
\begin{align}\label{Pf}
    P_f = V I,
\end{align}
and the actual power sent out by the battery is given by
\begin{align}\label{P}
    P = V I.
\end{align}
It is to be noted that expressions for both the power computed from sensor feedback $P_f$ and the actual power $P$ delivered by the battery are equal because we have not yet considered the effect of attacks. The P/u converter gives the command input to the vehicle $u$ by converting the delivered power $P$ as follows
\begin{align}\label{Ptou}
    u = \frac{1}{\kappa} P.
\end{align}

\subsection{CAEV Model under Cyber-Attacks}

The intention of the adversary is to cause vehicle collision between successive vehicles in the platoon. As mentioned earlier, we consider that the adversary has access to two attack surfaces on which it can create attack, (i) attack on DSRC network data $a_{c}$: $\Delta_a$, and (ii) attack on the current sensor data $I$: $\Delta_I$. In order to achieve its malicious objective the attacker can utilize any combination of these two attacks. We conisder that the adversary attacks both these surfaces simultaneously. This is considered because it covers both the special cases where the adversary attacks either the network data $a_{c}$ or the sensor data $I$. Thus giving the defender to build a secure control algorithm which will be effective against a larger set of attack vectors. Now, we discuss how the model equations of both the vehicle and battery are affected due to the attack vectors. Since the attack $\Delta_a$ is corrupting the acceleration data $a_c$ in intervehicular communication, its effect is directly observed in determining the required control input which is now given by
\begin{align}\label{ureqatt}
    u_{req}' = (k_p e + k_d \dot{e}) + a_c + \Delta_a.
\end{align}
This will send an incorrect power requirement to the battery. Similarly as the attack $\Delta_I$ is corrupting the current sensor data $I$, its effect appears directly in determining the feedback power for the current controller which is now given by
\begin{align}\label{Pfatt}
    P'_f = V (I + \Delta_I).
\end{align}
It is evident that the power computed from sensor feedback $P'_f$ and the actual power $P$ delivered by the battery are not equal now since there are attacks present unlike the nominal sccenario. This in turn affects the current circulating through the battery as determined by the BMS thus affecting the anode dynamics, the voltage output, the power delivered by the battery and eventually the control input commanded to the vehicle. Thus, the adversary can achieve its malicious objective of causing vehicle collision through coordinating between the attacks $\Delta_a$ and $\Delta_I$.

\section{Secure Control of CAEV}

In this section, we discuss the secure control approach adopted for CAEV.

\subsection{Overview of the CAEV Defense Approach}
We discussed the overall platoon operation earlier where the goal is to keep a desired intervehicular distance between successive vehicles while following CACC strategy where each vehicle is powered by an onboard battery which is in loop with the local vehicular controller thus resulting in a CAEV structure. As mentioned in the previous section, the adversary can create attacks by corrupting the intervehicular communication data and current sensor data. So, it is essential to have a secure controller which is robust against such attacks which we refer to as defender. The high level objective of the defender is to prevent the vehicles from colliding with each other under the presence of attacks. Specifically, the local vehicle controller must maintain the desired standstill intervehicular distance equal to $d_r$ at all times under the presence of attacks. The objective of the defender is stated mathematically as follow
\begin{align}
    e \rightarrow 0 \quad \text{under} \quad \Delta_a, \Delta_I \neq 0.
\end{align}

\begin{figure*}[t]
\begin{center}
\includegraphics[width=\textwidth]{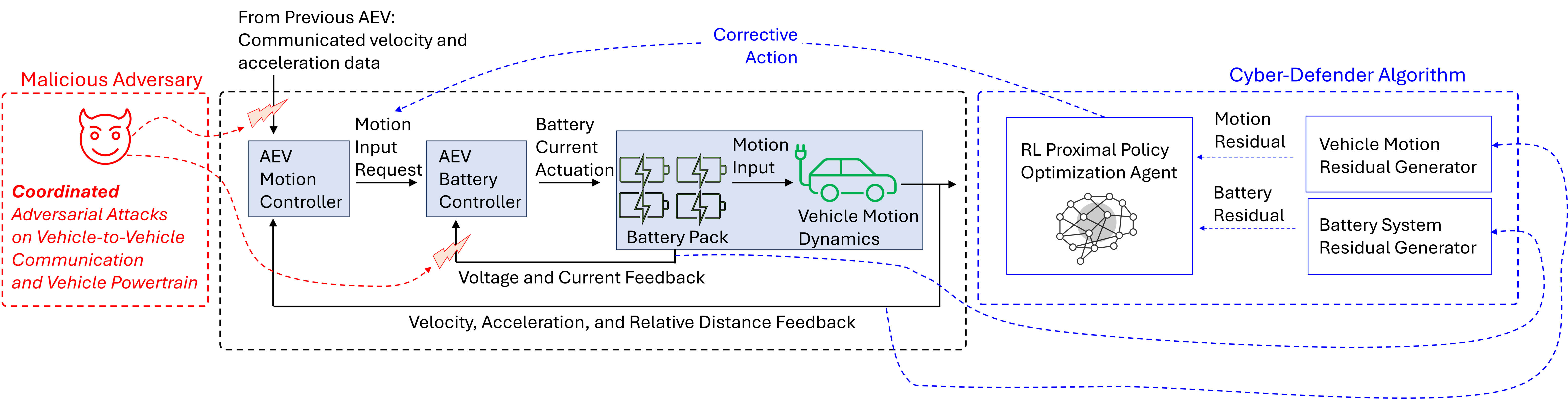}
\caption{A schematic of the proposed secure control framework for CAEV system using a reinforcement learning based cyber-defender algorithm.} 
\label{fig:diagControl}
\end{center}
\end{figure*}

The overall structure of the secure control architecture is shown in Fig. \ref{fig:diagControl}. The vehicle state measurements $d$, $w$ and $a$ from the sensors are sent to the local vehicular controller which is essentially an acceleration controller. It also receives the preceding vehicle's velocity data $v_c$ and the acceleration data with attack $(a_c+\Delta_a)$. Using this information it computes the required control input $u_{req}$. The u/P converter then sends the power requirement $P_{req}$ calculated using $u_{req}$ to the current controller of the BMS. The current controller receives the voltage data $V$ and the current data with attack $(I+\Delta_I)$ from the battery sensors. Using this information it calculates and circulates the current $I$ through the battery. Then the battery delivers the power $P$. This goes through the P/u converter which converts it to the acceleration input $u$ which is finally supplied to the powertrain. Due to presence of the acceleration attack $\Delta_a$ and the current attack $\Delta_I$, the input requirement $u_{req}$ and the actual input $u$ supplied to the vehicle powertrain will be different from what they would have been under nominal no attack scenario. In order to achieve our aforementioned secure control goal, the defender sends an additional control input $u_{RL}$ to the u/P converter. Qualitatively, $u_{RL}$ nullifies the effect of attacks $\Delta_a$ and $\Delta_I$ so that the battery powered CACC operation is safely undertaken without any vehicle collision even under the presence of earlier mentioned attacks.

\subsection{Details of the Defender Architecture}
Here, we discuss the defender architecture in detail. Specifically, we illustrate how the additional control input $u_{RL}$ is determined. The defender consists of three subsystems: (i) a vehicle motion residual generator, (ii) a battery system residual generator, and (iii) an RL agent. The residual generators are designed adopting a similar approach as used in \cite{firoozi2021cylindrical}. Here we briefly elaborate them for the sake of completeness.

The vehicle motion residual generator calculates an estimate $[\hat{d}, \hat{w}, \hat{a}]^T$ of the actual vehicle states $[d, w, a]^T$. In order to calculate the estimate it needs the velocity and acceleration data of the preceding vehicle. The estimator is equipped with the model of the preceding vehicle using which it calculates the $w_c$ and $a_c$. It is to be noted that this data is attack free since it neither transmitted through the communication network nor received through the local sensors. Considering the vehicle model \eqref{ssveh}-\eqref{err}, the state-space model model can be written as
\begin{align}
    & \dot{x}_v = A x_v + B u_v, \label{mot-model}
\end{align}
where $x_v = [d, w, a]^T$ consists of vehicle states and $u_v$ is the vehicle input consisting of $w_c$, $a_c$, and $d_r$ which comes from communication data and known control parameters. Based on \eqref{mot-est}, an estimator can be formed as:
\begin{align}
    & \dot{\hat{x}}_v = A \hat{x}_v + B \hat{u}_v + M_v ({x}_v - \hat{x}_v), \label{mot-est}
\end{align}
where $\hat{x}_v$ is the estimated state; ${x}_v$ comes from the sensor data; $\hat{u}_v$ is the estimated input computed using a model of the previous vehicle; and $M_v$ is the estimator gain. Defining the residual $r_v = {x}_v - \hat{x}_v$, and subtracting \eqref{mot-est} from \eqref{mot-model}, the residual dynamics can be written as:
\begin{align}
    & \dot{r}_v = (A-M_v) r_v + B \tilde{u}_v, \label{res-dyn}
\end{align}
where $\tilde{u}_v=u_v-\hat{u}_v$. Note that, under no attack conditions, $\tilde{u}_v=0$ as $u_v$ and $\hat{u}_v$ are almost identical. Under communication attack, the communicated data $u_v$ is corrupted by the attack while the locally computed data $\hat{u}_v$ is not corrupted, leading to $\tilde{u}_v \neq 0$. Here, $M_V$ is designed to ensure the eigen values $\lambda_v(A-M_v)\leq 0$ for stability purposes, and the transfer gain $\tilde{u}_v$ to $r_v$ has desired magnitudes for residual sensitivity purposes. A similar approach was taken to design the battery system residual generator which outputs the residual $r_b$.

The residuals $r_v$ and $r_b$ are sent to the RL agent. If the residual $r_v = 0$ the agent takes the action $u_{RL}=0$ otherwise it takes some action $u_{RL}$ determined by its policy. We have used the Proximal Policy Optimization (PPO) RL agent \cite{schulman2017proximal}. The environment for the agent is the entire vehicle and battery integrated ecosystem which corresponds to all the blocks in Fig. \ref{fig:diagControl} except the block labeled `Defender: RL Proximal Policy Optimization Agent'. From this environment the agent receives the measurements $[d, w, a]^T$. Using this signals the agent calculates the error $e$ as defined in Section II. We define the observation space for the RL agent as the 2 element vector $[e, r_v]^T$. The action space from which the action $u_{RL}$ is sampled is continuous real scalar field. The action space is chosen to be bounded since there are physical limits on the control input that can be supplied to the vehicle powertrain. The bounds are $a_{min} \leq u_{RL} \leq a_{max}$ . The reward function for the agent is chosen such that the agent learns to take action 
$u_{RL}$ in order to maintain the defender's objective of $e \rightarrow 0$ under the presence of attacks $\Delta_a$ and $\Delta_I$. Specifically, we give a positive reward to the agent if it is able to achieve the defender's objective within certain threshold and we penalize the agent for making mistakes, i.e., not achieving the defender's objective. Mathematical description of the reward model is as follows
\begin{align}\label{rewardfn}
    \text{reward} = 
        \begin{cases}
            \quad 10, \quad & |e| \leq 1\\
            -1000, \quad & \text{otherwise}.
        \end{cases}
\end{align}
PPO agent learning process involves training the agent. The training algorithm belongs to a class of policy gradient methods. The objective function used by the PPO agent is clipped to maintain training stability and it incorporates a trust region to limit drastic changes in the policy between successive updates. These features enhance PPO's robustness to RL environment and make it suitable for our task since we want to capture the attack scenarios which contribute to the unwanted changes in the environment.

%%%%%%%%%%%%%%%%%%%%%%%%%%%%%%%%%%%%%%%%%%%%%%%%%%%%%%%%%%%%%%%%%%%%%%%%%%%%%%%%
\section{Results and Discussion}

\begin{figure}
\begin{center}
\includegraphics[width=0.48\textwidth]{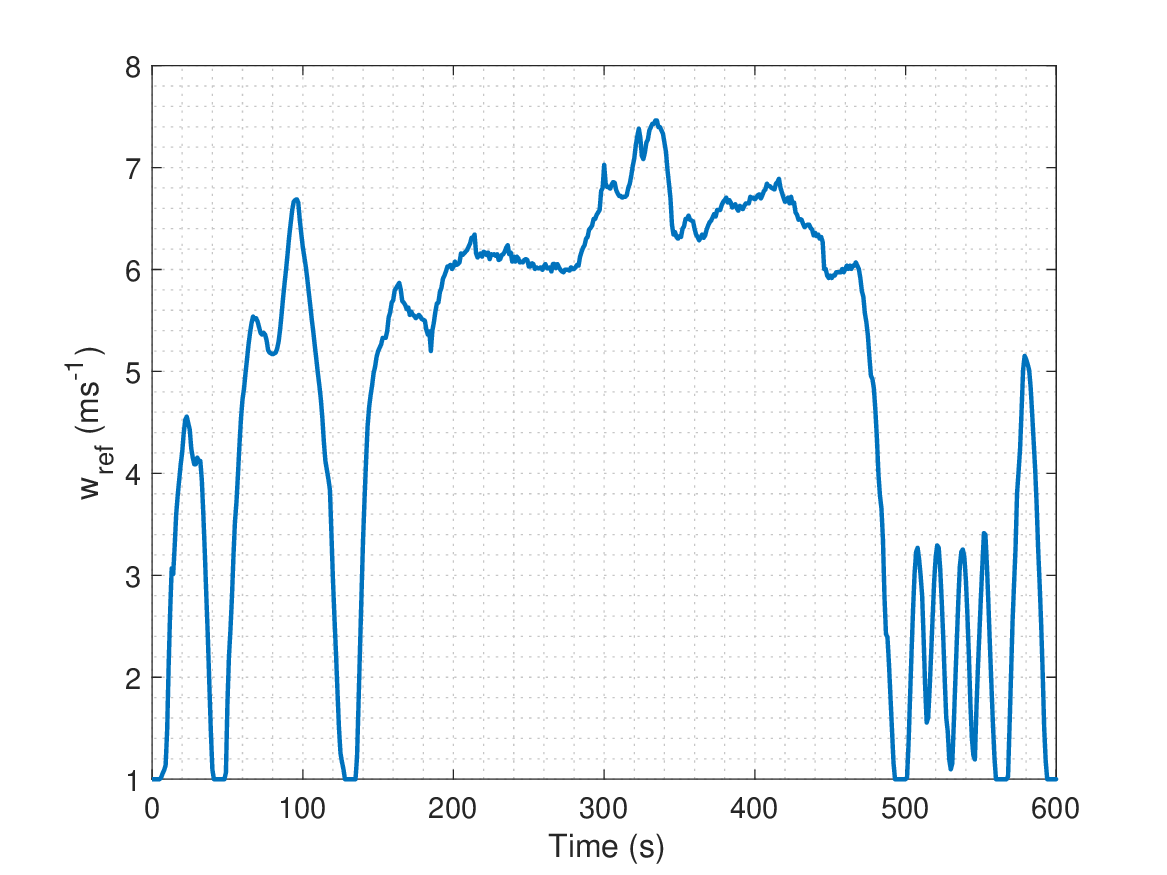}
\caption{Reference velocity trajectory for leader vehicle.}
\label{fig:vref}
\end{center}
\end{figure}

In this section, we present the case study results from simulations to illustrate the performance of RL PPO agent in defending against the adversarial cyber attacks. In the simulations we have considered the vehicle immediately following the leader vehicle to be our ego vehicle. We have used MATLAB and Simulink environments for training the agent and simulating the CAEV model. The reference velocity profile followed by the leader vehicle is based on modified Supplemental Federal Test Procedure (US06) as shown in Fig. \ref{fig:vref}. Battery model parameters are as per \cite{firoozi2021cylindrical} and the vehicle local controller parameters are as per \cite{biron2018real}. All of the required simulation parameters are listed in Table. \ref{tb:simparam}.

\begin{table}[hb]
\begin{center}
\caption{Simulation parameters}\label{tb:simparam}
\begin{tabular}{cc|cc}
Parameter & Value & Parameter & Value \\ \hline
$h$ & $0.6s$ & $\kappa$ & $10Wm^{-1}s^{2}$ \\
$k_p$ & $0.7s^{-2}$ & $K_b$ & $0.1AW^{-1}$ \\
$k_d$ & $1.0s^{-1}$ & $a_{min}$ & $-10ms^{-2}$ \\
$d_r$ & $5m$ & $a_{max}$ & $10ms^{-2}$ \\ \hline
\end{tabular}
\end{center}
\end{table}

\begin{figure}
\begin{center}
\includegraphics[width=0.48\textwidth]{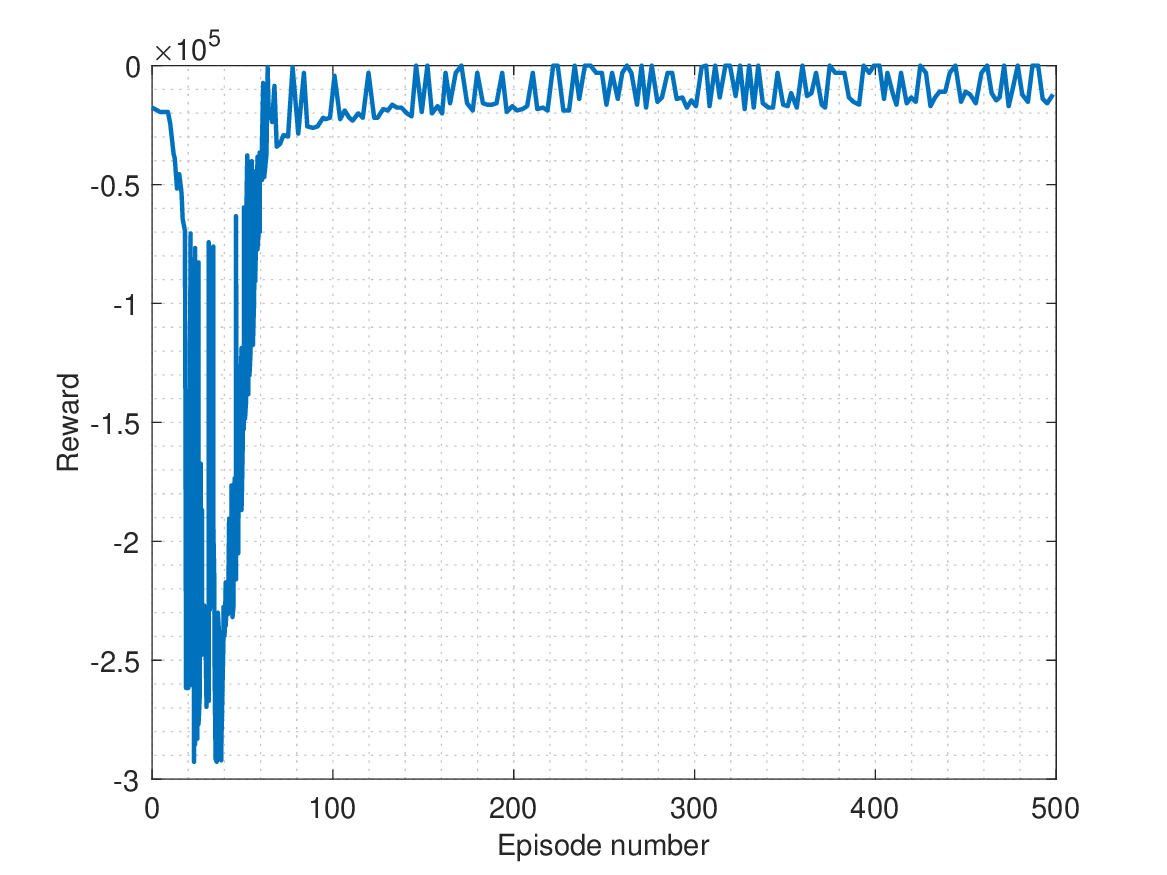}
\caption{Training progress of the PPO agent: rewards vs episode.}
\label{fig:reward}
\end{center}
\end{figure}

We create a custom environment defining the observation and action spaces as well as the reward function as mentioned in Section III to train the PPO agent. Specifically, we define the environment and agent `objects' in MATLAB to be used by the `RL agent block' in the Simulink model and define the reward function using Simulink model signals. The agent is trained for a maximum number of $500$ episodes and the maximum number of timesteps chosen per episode is $300$. The agent is trained under a variety of attack scenarios, i.e., using numerous combination of $\Delta_a$ and $\Delta_I$. The reward learned during training process is shown in Fig. \ref{fig:reward}. It can be seen that the reward received steadily increases at the start and then stablizes which indicates that the agent quickly learns to take actions to achieve the defender objective. This pretrained agent is saved and then deployed under adversarial CAEV operation. It is to be noted that the agent does not have the knowledge of the attack vectors when it is being implemented online.

% \textcolor{red}{discussion on nominal scenario. Show velocity, relative distance, accl input, and battery current, SOC and voltage plots.}
\subsection{Case Study 1: Agent Performance under Coordinated Attacks}
\begin{figure}
\begin{center}
\includegraphics[width=0.48\textwidth]{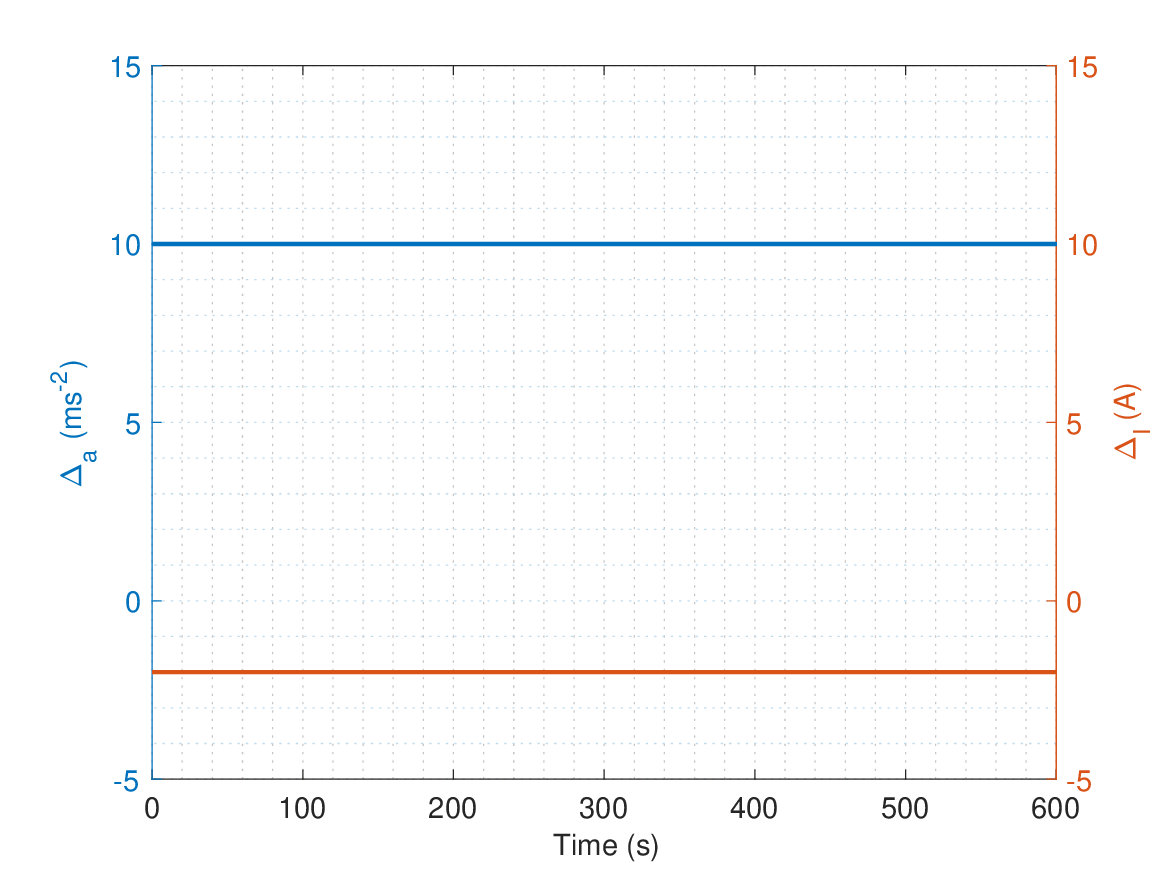}
\caption{Adversarial vector profiles: Acceleration attack, and Current attack.} 
\label{fig:attvec}
\end{center}
\end{figure}

The adversarial attack vectors used for simulations in this case study are shown in Fig. \ref{fig:attvec}. The attack vectors are chosen to replicate a typical attack studied in literature, namely, False Data Injection (FDI) attack. FDI infuses falisified information in the data streams. Here, we consider step input attacks of magnitude $\Delta_a = 10ms^{-2}$ in the communication network and $\Delta_I=-2A$ on the current sensor. These attacks are injected at the start of the simulation, i.e., at $t=0s$.

\begin{figure}
\begin{center}
\includegraphics[width=0.48\textwidth]{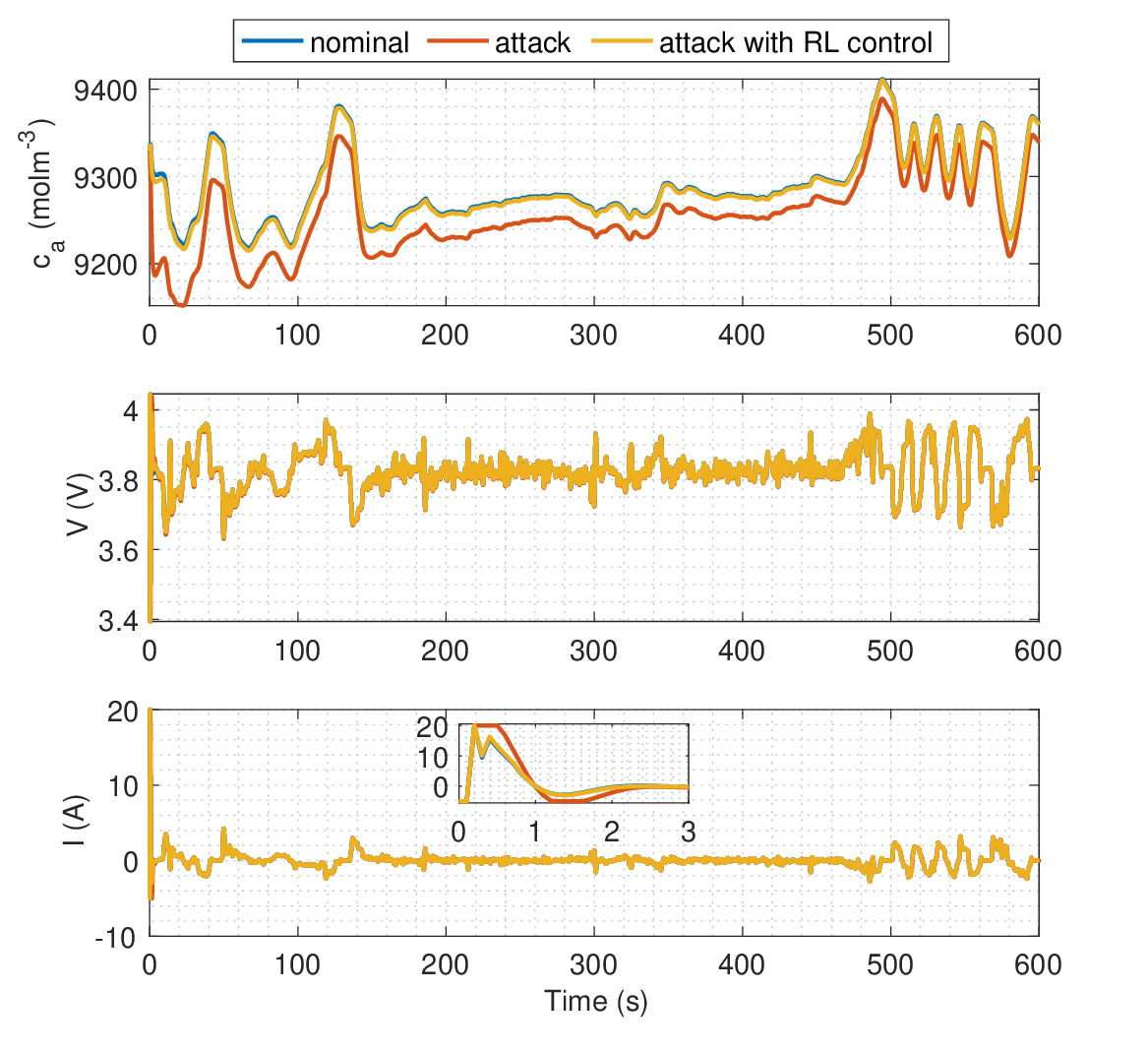}
\caption{Variation of battery data: (Top) Anode concentrations (Middle) Voltage, and (Bottom) Current.} 
\label{fig:cVI}
\end{center}
\end{figure}

The anode surface concentration, voltage and current in the battery under various scenarios are shown in Fig. \ref{fig:cVI}. It can be seen from the bottom plot of Fig. \ref{fig:cVI} that at the onset of attacks the current demand rises rapidly. The effect of this abnormal current requirement is seen in the surface anode concentration. This can be seen in the top plot of Fig. \ref{fig:cVI} where the anode concentration is decreased rapidly under the attack scenario as shown by the red line. Whereas the RL agent is able to keep the concentration levels same as in nominal conditions of no attack, thereby shown the potency of the defender.

\begin{figure}
\begin{center}
\includegraphics[width=0.48\textwidth]{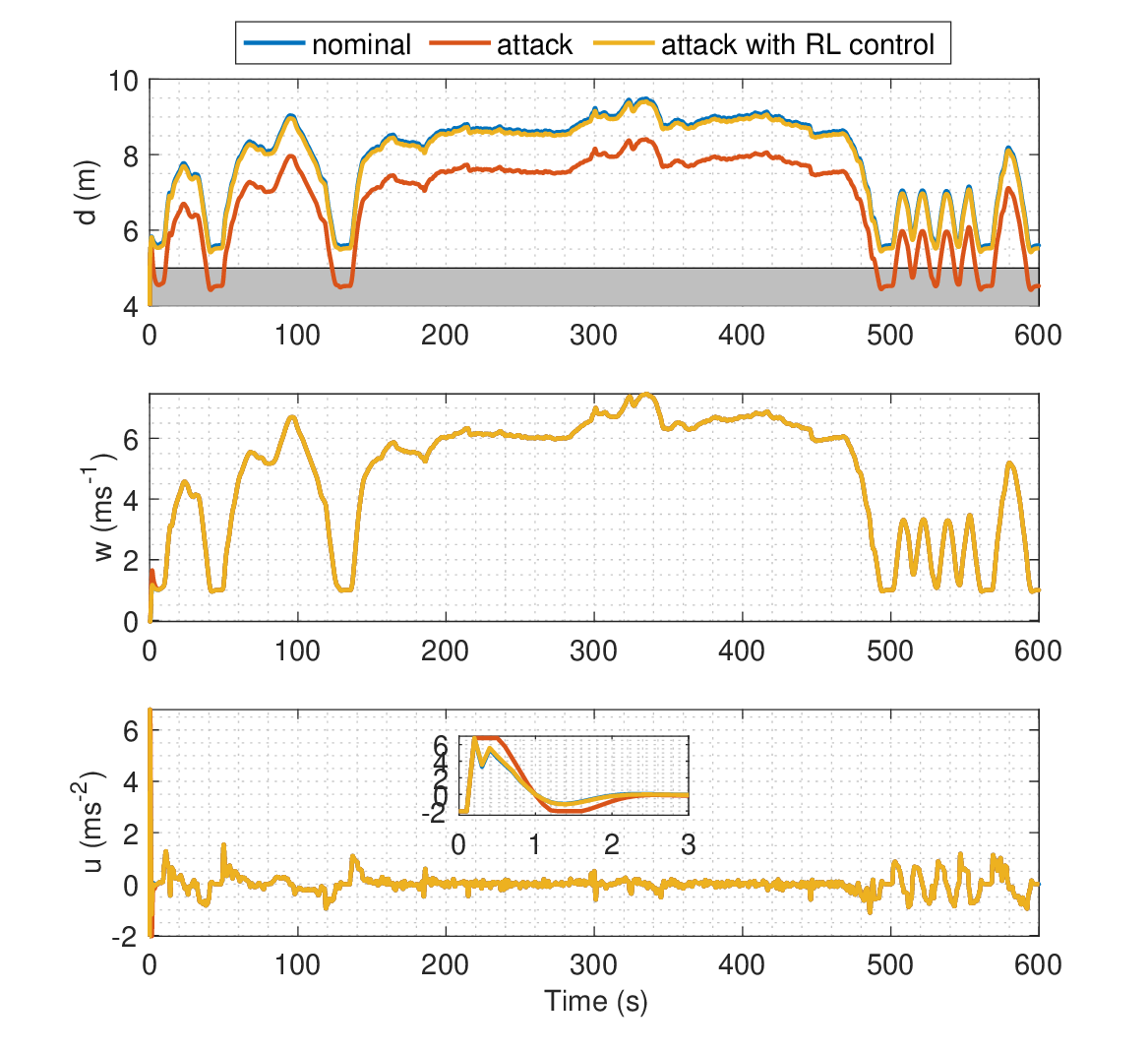}
\caption{Variation of vehicle data: (Top) Intervehicular distance, (Middle) Velocity, and (Bottom) Control input to powertrain.} 
\label{fig:dwu}
\end{center}
\end{figure}

The intervehicular distance between the ego vehicle and the leader, ego vehicle's velocity and acceleration under various scenarios are shown in Fig. \ref{fig:dwu}. The grey shaded portion in the top plot of Fig. \ref{fig:dwu} is the unsafe region, i.e., where the standstill intervehicular distance is less than $d_r$ eventually leading to vehicle collision. It is evident that the ego vehicle enters the unsafe region when there are attacks. But the RL agent is indeed able to defend against the attacks as shown by the yellow line which indicates that the agent keeps safe intervehuicular distance similar to the nominal scenario (shown by blue line). As compared to the nominal scenrio, the control input to the powertrain follows a slightly different path under attack scenario as shown by the red line in the bottom plot of Fig. \ref{fig:dwu}. Still, this variation at the start is enough to safety issues as discussed earlier. However, the yellow line in the top plot of Fig. \ref{fig:dwu} shows the effectiveness of our PPO agent in keeping the intervehicle distance safe thus preventing the collisions and validating the performance of the trained agent under adversarial conditions.

% \textcolor{red}{discussion on attacked scenario with no control. Show velocity, relative distance, accl input, and battery current, SOC and voltage plots.}

% \textcolor{red}{discussion on attacked scenario with your control approach. Show velocity, relative distance, accl input, and battery current, SOC and voltage plots.}

\begin{figure}
\begin{center}
\includegraphics[width=0.48\textwidth]{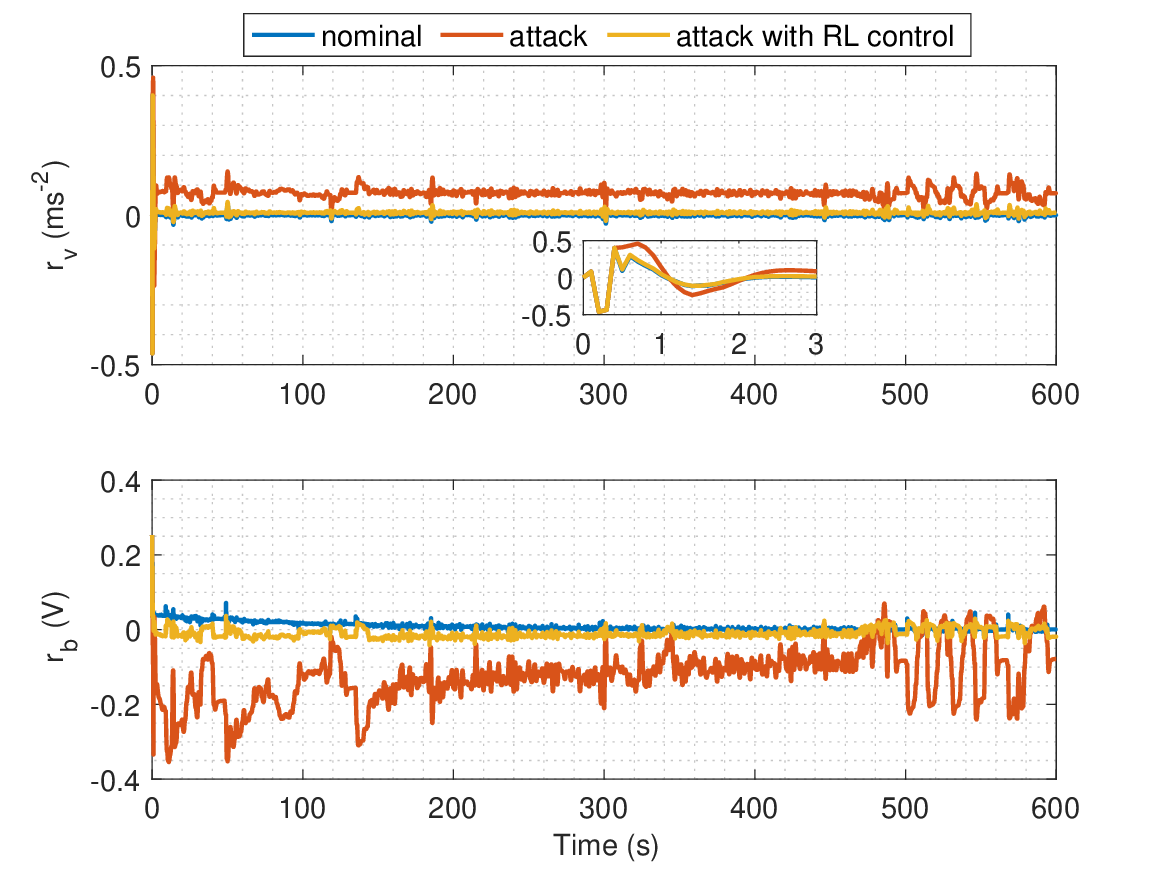}
\caption{Residuals: (Top) Vehicle motion residual, and (Bottom) battery system residual.} 
\label{fig:residual}
\end{center}
\end{figure}

\begin{figure}
\begin{center}
\includegraphics[width=0.48\textwidth]{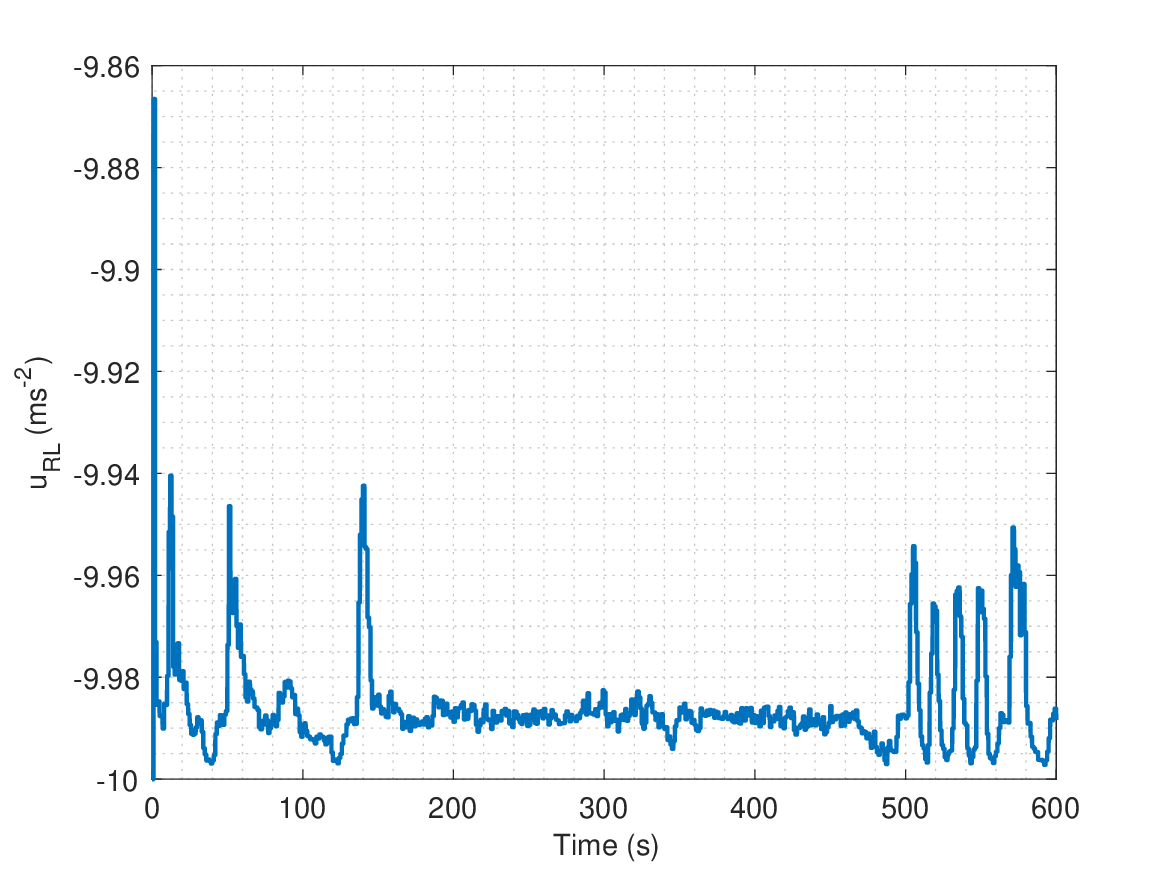}
\caption{Input action taken by the PPO agent.} 
\label{fig:uRL}
\end{center}
\end{figure}

The residuals are a key part of the defender algorithm which are shown in Fig. \ref{fig:residual}. Both the residuals $r_v$ and $r_b$ serve as the attack detector signals to the RL agent. Under nominal scenario both the residuals are equal to zero which is expected since the estimator gains are chosen such that the estimator has stable dynamics. Under the presence of attacks the residuals do not converge to zero as shown by the red lines in Fig. \ref{fig:residual}. As soon as the RL agent detects this change it sends the additional control input $u_{RL}$ as shown in Fig. \ref{fig:uRL} to nullify the effect of the attacks. The agent action $u_{RL}$ is able to stabilize the estimator dynamics to zero as shown by the yellow lines in Fig. \ref{fig:residual} which is the same as in nominal scenario. Further the magnitude of $u_{RL}$ is within the bounds of the action space indicating the feasibility of the additional control input to the powertrain. Hence, the RL PPO agent acting as the defender under adversarial cyber attacks is successful in mitigating the attacks thus preventing vehicle collision by maintaining safe intervehicluar distance while giving feasible actions.

\subsection{Case Study 2: Agent Performance under Limiting Attack Scenarios}
In this case study, we focus on the limiting cases of attack vectors. Specifically, we evaluate the maximum magnitudes of the attack vectors, $\Delta_a$ and $\Delta_I$, under which the PPO agent is able to keep the desired intervehicular distance.

\begin{figure}
\begin{center}
\includegraphics[width=0.48\textwidth]{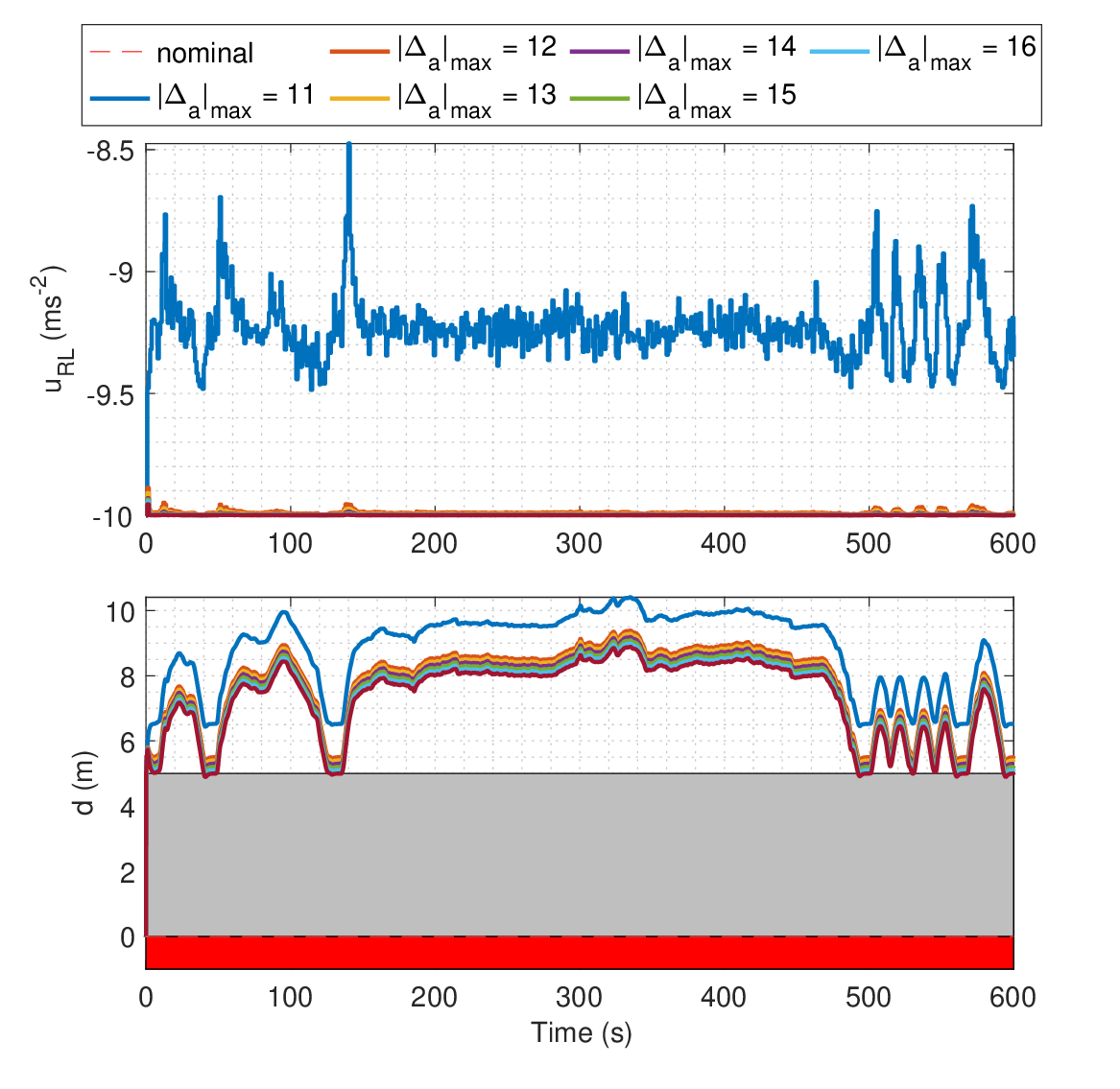}
\caption{(Top) Agent action taken, and (Bottom) Intervehicular distance, under different acceleration attack vectors.}
\label{fig:cs2a}
\end{center}
\end{figure}

First, we discuss the effect of increasing the maximum value $|\Delta_a|_{max}$ of the acceleration attack magnitude on the performance of the agent. The attack value is increased from $11 ms^{-2}$ to $16 ms^{-2}$ which leads to variation in the agent's control action as shown in the top plot of Fig. \ref{fig:cs2a}. It can be seen that the agent takes the action $u_{RL}$ in order to negate the effect of the maximum attack magnitudes. As the attack magnitude increases $u_{RL}$ gets saturated at $-10 ms^{-2}$ since that is the lower bound on $a_{min}$ on $u_{RL}$. The effect of the agent's command is reflected in the intervehicular distance $d$ whether it is less than the desired safety spacing $d_{des}$. As shown in the bottom plot of \ref{fig:cs2a}, the agent is able to maintain safe operation, i.e., $d \geq d_{des}$, till the maximum attack value reaches $15 ms^{-2}$. Beyond which $d$ becomes less than $d_{des}$. Hence, we conclude that the agent can mitigate acceleration attack vectors upto a maximum value of $15 ms^{-2}$.

\begin{figure}
\begin{center}
\includegraphics[width=0.48\textwidth]{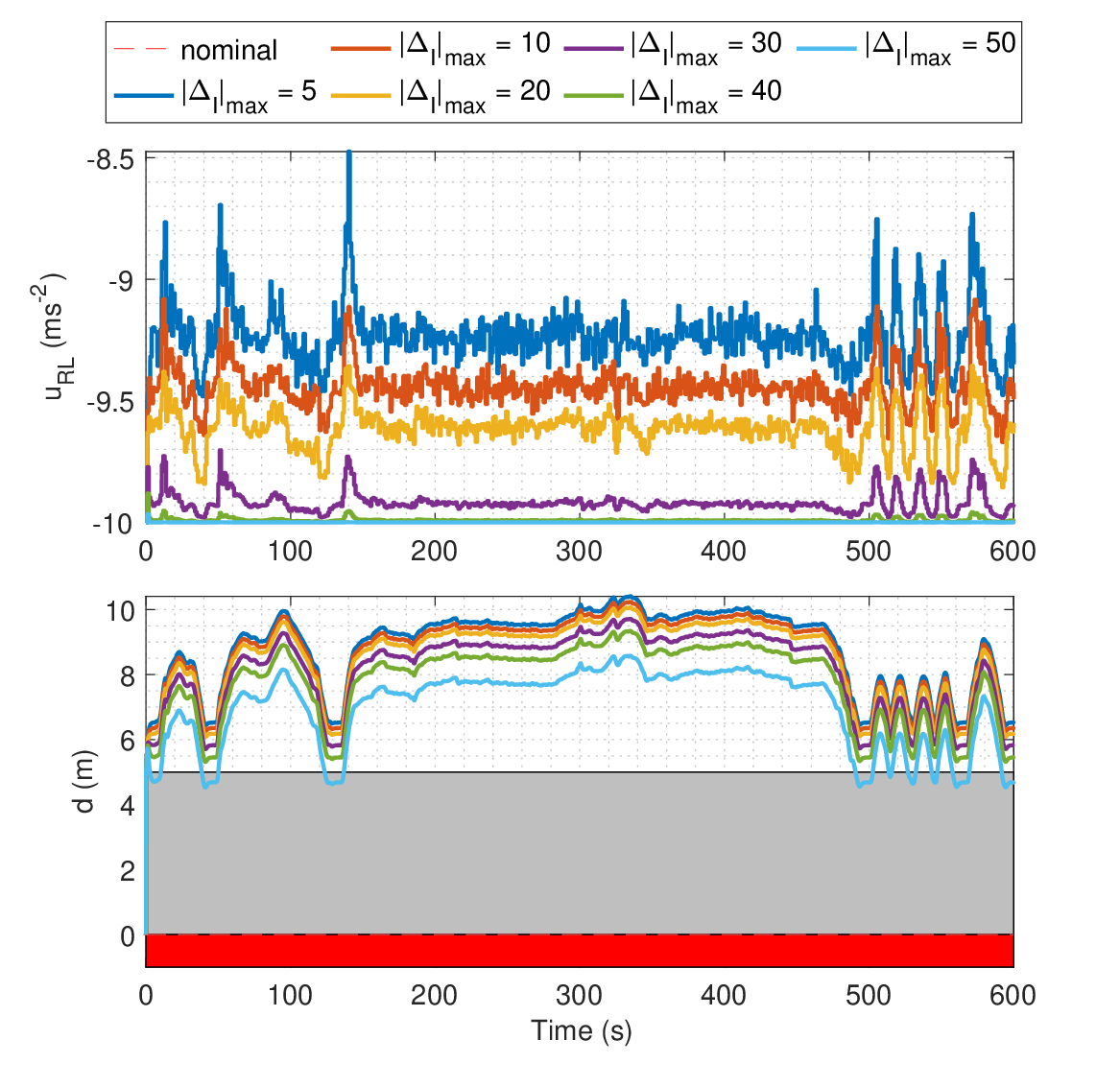}
\caption{(Top) Agent action taken, and (Bottom) Intervehicular distance, under different current attack vectors.}
\label{fig:cs2I}
\end{center}
\end{figure}

Next, we analyze the effect of incrementing the maximum value $|\Delta_I|_{max}$ of the current attack magnitude on the performance of the agent. The attack value is changed from $5 A$ to $50 A$ which leads to change in the agent's control action as shown in the top plot of Fig. \ref{fig:cs2I}. It can be seen that the agent takes the action $u_{RL}$ so that the effect of the maximum attack magnitudes is nullified. With the attack magnitude being increased, $u_{RL}$ becomes higher in magnitude eventually getting clipped at lower bound at $a_{min}$. The impact of the agent's response is evident in the evolution of the intervehicular distance $d$ with respect to the desired safety spacing $d_{des}$. The bottom plot of \ref{fig:cs2a} shows that the agent is able to keep the vehicles from colliding till the maximum attack value reaches $40 A$. After which $d$ enters the unsafe region. Thus, we conclude that the agent can suppress current attack vectors reaching to a magnitude of $40 A$.

%%%%%%%%%%%%%%%%%%%%%%%%%%%%%%%%%%%%%%%%%%%%%%%%%%%%%%%%%%%%%%%%%%%%%%%%%%%%%%%%
\section{Conclusions}
In this paper, we proposed a secure control architecture in the presence of adversarial cyber attacks in CAEV. Specifically, we designed a defender using an RL agent which provides an additional control input along with the input supplied by the battery to the vehicle powertrain. Through simulation case studies, we effectively illustrated the performance of the RL agent to defend against the attacks on current sensor and acceleration information. The effectiveness of the RL agent in the limiting attack scenarios is evaluated by determining the maximum attack vectors the agent is able to mitigate. For future work, we intend to incorporate attacks on local controller parameters, consider coordinated velocity and acceleration communication network attacks as well as voltage and current sensor attacks, consider multi-vehicle attack cases and further explore platoon congestion scenario.

\bibliographystyle{IEEEtran}
\bibliography{ref.bib}

% \begin{IEEEbiography}[{\includegraphics[width=1in,height=1.25in,clip,keepaspectratio]{figures/photo_Shashank.jpg}}]{Shashank Dhananjay Vyas}
% received a dual B.Tech. + M.Tech in Mechanical Engineering from IIT Kharagpur, India in 2019. Later worked at Bajaj Automotive Ltd, India. Currently he is a Ph.D. candidate at The Pennsylvania State University working on Connected Autonomous Vehicles (CAVs). His primary research interests are optimisation, controls and machine learning in the realm of CAVs.
% \end{IEEEbiography}

% %\vspace{-43pt}

% \begin{IEEEbiography}[{\includegraphics[width=1in,height=1.25in,clip,keepaspectratio]{figures/Dey.jpg}}]{Satadru Dey}
% (Senior Member, IEEE) received the master’s degree in control systems from the Indian Institute of Technology Kharagpur, Kharagpur, India, in 2010, and the Ph.D. degree in automotive engineering from Clemson University, Clemson, SC, USA, in 2015.
% He is an Assistant Professor with the Department of Mechanical Engineering, The Pennsylvania State University, University Park, PA, USA. From August 2017 to August 2020, he was an Assistant Professor with the University of Colorado Denver, Denver, CO, USA. He was a Postdoctoral Researcher with the University of California at Berkeley, Berkeley, CA, USA, from 2015 to 2017. His technical background is in the area of controls and his research interest lies in smart cities, energy, and transportation systems.
% \end{IEEEbiography}

\end{document}